\documentclass[12pt, aps, prb, preprint]{revtex4}
\usepackage{graphicx}
\usepackage{epsfig}
\usepackage{latexsym}
\usepackage{amssymb}
\usepackage{bm}
\usepackage{float}
\usepackage[usenames]{color}
\usepackage{subfigure}
\usepackage{amsmath}
\usepackage{amsfonts}

\begin{document}

\title{
An analysis of the LIGO discovery based on Introductory Physics} 
\author{Harsh Mathur} 
\affiliation{Department of Physics, Case Western Reserve University, Cleveland, Ohio 44106-7079}
\author{Katherine Brown}
\author{Ashton Lowenstein}
\affiliation{Department of Physics, Hamilton College, Clinton, NY 13323 }

\date{\today}

\begin{abstract}

By observing gravitational radiation from a binary black hole merger, the
LIGO collaboration has simultaneously opened a new window on the universe and 
achieved the first direct detection of gravitational waves. Here this discovery is
analyzed using concepts from introductory physics. 
Drawing upon Newtonian mechanics, dimensional considerations, and analogies
between gravitational and electromagnetic waves, we are able to explain the principal
features of LIGO's data and make order of magnitude estimates of key parameters of the 
event by inspection of the data. Our estimates of the black hole masses, the distance to the event,
the initial separation of the pair, and the stupendous total amount of energy radiated are all in 
good agreement with the best fit values of these parameters obtained by the LIGO-VIRGO collaboration. 


\end{abstract}

\maketitle

\section{Introduction}

On September 14, 2015 the LIGO collaboration detected gravitational radiation from the
merger of a binary black hole a billion light years away.\cite{ligo} This discovery is a major scientific breakthrough
that constitutes the first direct observation of gravitational radiation almost a hundred years after 
Einstein predicted it. It has also opened a new window on the Universe.\cite{viewpoints} 
Arguably the LIGO observation is the most sensitive
measurement ever made in the history of science. The experiment has generated tremendous
interest both amongst the general populace as well as amongst students taking physics. Thus it
represents an exceptional pedagogical opportunity.

The event observed by LIGO (designated GW150914) consisted of the merger of two black holes with masses 
equal to 29 and 36 solar masses, respectively. The process took approximately a tenth of a second.
The energy released in the form of gravitational radiation was the energy equivalent of three solar masses.
The merger took place at a distance of a billion light years and hence a billion years ago.
The direction of the source was only partially determined by the two LIGO detectors. All of these facts were inferred
from data summarized in figure 1 (reproduced from the discovery paper by the LIGO collaboration). \cite{ligo}
The parameters quoted were extracted by fitting the data to templates generated by state of the
art numerical relativity. 

The purpose of this article is to explain key features of the data in figure 1 in terms of
introductory physics and to use figure 1 to make back of the envelope estimates of the
parameters quoted in the preceding paragraph. We make simple arguments based on
Newtonian gravity, dimensional analysis and a rudimentary acquaintance with electromagnetism 
and waves. 
Black holes and their merger are relativistic phenomena and the peak 
of the observed signal comes from the coalescence of the two black holes, which is
highly relativistic. To fully understand the underlying physics of a binary black hole
merger in general and figure 1 in particular requires both analytic and numerical 
general relativity. The Newtonian treatment given here can only provide an 
understanding at an order of magnitude level. Our treatment is aimed at students
in introductory college level physics courses, as well as high school students 
taking AP or IB Physics, who do not have a preparation in general relativity 
but want to understand the LIGO result.


\begin{figure*}
\includegraphics[width=\textwidth]{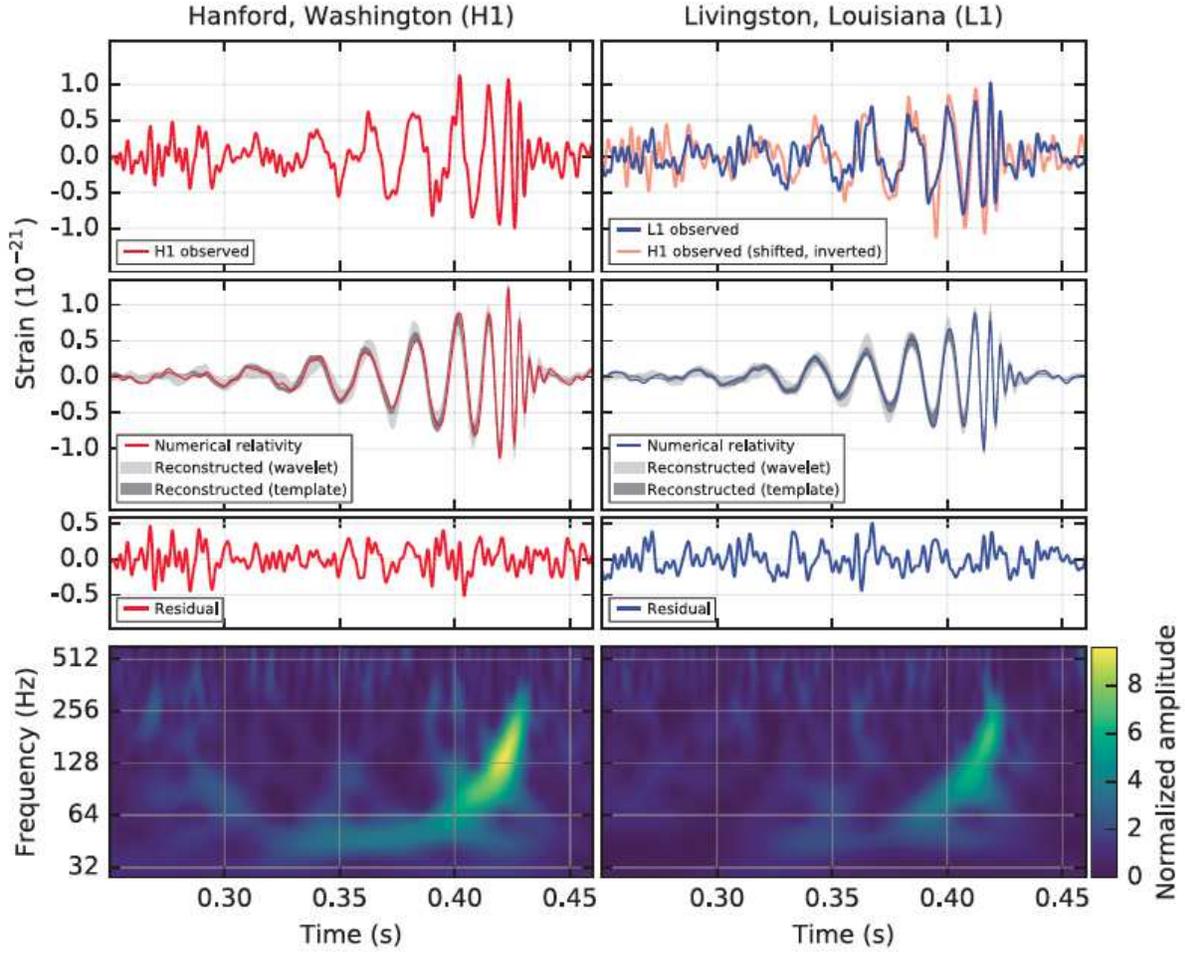}
\caption{Summary of LIGO data (reproduced from the discovery paper).$^{ \cite{ligo}}$ The top left panel shows the strain $h$ observed by the Hanford detector as a function of time; the top right panel shows the data for the Livingston detector with the Hanford data time-shifted, inverted and overlaid to show the excellent match between the two detectors. The data have been band pass filtered to lie in the 35-350 Hz band of maximum detector sensitivity; spectral line noise features in the detectors within this band have also been filtered. The second row shows a fit to the data using sine-Gaussian wavelets (light gray) and a different waveform reconstruction (dark gray). Also shown in color are the signals obtained from numerical relativity using the best fit parameters to the data. The third row shows for both detectors  the residuals obtained by subtracting the numerical relativity curve from the filtered data in the first row. The fourth row gives a time-frequency representation of the data and shows the signal frequency increasing in time.}
\end{figure*}

Binary black hole mergers take place in three stages. Initially the black holes circle their common
center of mass in essentially circular orbits. During this stage they lose orbital energy in the form
of gravitational radiation and spiral inward. In the second stage the black holes coalesce to form
a single black hole. In the third stage, called ring-down, the merged object relaxes into its equilibrium
state called a Kerr black hole.\cite{kerr} Gravitational radiation is emitted copiously during merger and ring-down
as well but it is the in-spiral stage that is conducive to simple analysis, and that
is the basis of the back of the envelope estimates presented here. By contrast the merger and ring-down
are decidedly less simple. 
The ring down process
is not a standard component of introductory courses or textbooks on general
relativity but it can be presented at that level.\cite{chandrasekhar}
The merger presents a formidable problem in numerical relativity that had until recently resisted all attempts at solution.\cite{numericalreview}

The remainder of this paper is organized as follows. 
In section II we use simple arguments based on introductory physics to elucidate the underlying
physics, and we use the data shown in figure 1 to estimate the masses of the black holes,
their distance from earth, their initial separation and the total gravitational energy radiated. 
Along the way we define the chirp mass of a binary black hole system and derive the only equation that appears  
in the discovery paper.\cite{ligo}
We also present a lower bound on the total 
mass of the black hole pair.
In our analysis we do
not estimate the rotational angular momentum of the black holes either before or after the merger. 
In section III we discuss this and other limitations of our analysis as well as other matters of
pedagogical interest. In order to make our article more useful for instructors as well as for self study
we provide some problems in appendix \ref{sec:probs}.

\section{Parameter estimation} 

{\em Kepler's third law.} During the in-spiral the two black holes can be modeled as point masses
of mass $M$ and $m$ that move around their common center of mass in circular orbits of radii
$R$ and $r$ respectively at a common angular frequency $\omega$. The black holes remain
on opposite sides of the center of mass and hence their separation is $R + r$. Obviously
$Mr = mR$. At least initially, when the
motion is slow and the black holes are well separated,
Newtonian mechanics and Newton's law of gravitation should apply. It is then 
easy to show that the frequency is related to the separation of the two black holes via
\begin{equation}
\omega^2 = \frac{G(M+m)}{(R + r)^3}, 
\label{eq:keplerlaw}
\end{equation}
which is one form of Kepler's third law. 
We consider circular orbits
for simplicity since our goal is only to make rough estimates. More careful
analysis shows that even if the orbits were initially elliptical then emission of gravitational
radiation will in any case quickly circularize the orbits.\cite{peters}


{\em Orbital energy.} Continuing the Newtonian analysis of the orbiting black holes it is
useful to calculate the total energy of the system, namely, the sum of the orbital kinetic 
energy of both black holes and their mutual gravitational potential energy.\cite{ftnt:selfenergy}
A short calculation reveals \cite{ftnt:energy}
\begin{equation}
E_{{\rm tot}} = - \frac{1}{2} \frac{GMm}{(r+R)}.
\label{eq:energy}
\end{equation}
Eq (\ref{eq:energy}) gives the energy of the
system as a function of the black hole separation. For later use it is convenient to
use eq (\ref{eq:keplerlaw}) to express the energy as a function of the frequency of the
orbit instead.
\begin{equation}
E_{{\rm tot}} = - \frac{1}{2} \frac{ G^{2/3} M m}{(M+m)^{1/3}} \omega^{2/3}.
\label{eq:etotfreq}
\end{equation}

{\em Moment of inertia.} Another useful preliminary result is to calculate the moment of inertia 
of the orbiting black holes about their common center of mass, and to 
express the
result in terms of the masses of the black holes and their separation $(R+r)$. The result is
\begin{equation}
{\cal I} = \frac{ m M}{m + M} (r + R)^2.
\label{eq:moment}
\end{equation}
Note that as the distance between the black holes changes, 
the moment of inertia changes too, so it is not constant in time.

{\em Gravitational radiation emission.} 
Within the Newtonian framework there is no 
gravitational radiation and the circular orbit analyzed above should persist forever.
However in general relativity the orbiting black holes will emit gravitational radiation
thereby losing energy and spiraling towards each other. In order to analyze this process
we need to develop a formula for the rate of gravitational radiation emission. 
Gravitational radiation is produced by the motion of
large massive objects much as electromagnetic radiation	is
produced by the accelerated motion of charged particles.
Under suitable conditions the electromagnetic radiation	is
determined by the changing electric dipole moment of the
charge distribution. Similarly under suitable conditions
the gravitational radiation is produced by the changing
quadrupole moment of the mass distribution; the quadrupole
moment is essentially the same things as the moment of	
inertia which is familiar from introductory physics.\cite{ftnt:dipolerad}
By analogy to electromagnetic dipole radiation we expect that the
gravitational radiation field should be	proportional to	the quadrupole
moment that sources it and hence the radiated power
$P_{{\rm rad}} \propto {\cal I}^2$. This is becasue the energy
density in a gravitational wave	depends on the square of the field
much as the energy density in an electromagnetic wave depends on
the squares of the electric and	magnetic fields.
We also expect the radiated
power to depend on the frequency $\omega$ at which the radiating system is oscillating and on
the fundamental constants $G$ and $c$. Hence on dimensional grounds we expect
\begin{equation}
P_{{\rm rad}} = \alpha {\cal I}^2 \omega^{\xi} G^{\eta} c^{\zeta}.
\label{eq:dimensional}
\end{equation}
Here $\alpha$ is a dimensionless numerical factor and $\xi$, $\eta$ and $\zeta$ are exponents
that can be determined by dimensional analysis. It is easy to verify that
\begin{equation}
P_{{\rm rad}} = \alpha \frac{G {\cal I}^2 \omega^6}{c^5}.
\label{eq:quadrupolepower}
\end{equation}
The constant $\alpha$ cannot be as simply determined. A full analysis based on general relativity 
is needed and reveals that $\alpha = 32/5$. \cite{weinberg} Finally we note that to the extent that the
quadrupole approximation itself is valid one can show that the angular frequency of the radiation 
emitted is $2 \omega$. The reason for the factor of two is explained in appendix
\ref{sec:quadrupole} where a more rigorous version of eq (\ref{eq:quadrupolepower}) is also provided. 
Appendix \ref{sec:quadrupole} requires some familiarity with the moment of inertia tensor which is not
a standard component of an introductory physics course; however the appendix is outside
the main line of development of this article and can be safely skipped. \cite{feynmanvol2}

{\em Energy Balance.} Making use of eqs (\ref{eq:quadrupolepower}) and (\ref{eq:moment}) and
(\ref{eq:keplerlaw}) we see the power of gravitational radiation emission by the binary black holes
is given by
\begin{equation}
P_{{\rm binary}} = \alpha \frac{G^{7/3} \omega^{10/3}}{c^5} \frac{m^2 M^2}{(m+M)^{2/3}}
\label{eq:binarypower}
\end{equation}
On the other hand the rate at which the binary black holes lose orbital energy is obtained by differentiating
eq (\ref{eq:etotfreq})
\begin{equation}
- \frac{dE_{{\rm tot}}}{dt} = \frac{1}{3} G^{2/3} \frac{Mm}{(M+m)^{1/3}} \omega^{-1/3} \frac{d \omega}{dt}.
\label{eq:energyloss}
\end{equation}
Equating the energy loss (\ref{eq:energyloss}) to the power radiated (\ref{eq:binarypower}) yields 
\begin{equation}
\frac{ ( m M )^{3/5}}{(m+M)^{1/5}}= \frac{c^3}{G} \left( \frac{1}{3 \alpha} \omega^{-11/3} \frac{d \omega}{d t} \right)^{3/5}.
\label{eq:energybal}
\end{equation}
It is conventional to define the chirp mass ${\cal M}$ as the left hand side of eq (\ref{eq:energybal}). 
The chirp mass is a crucial scale in the in-spiral process. 

{\em The Chirp.} In order to compare to the data it is helpful to rewrite eq (\ref{eq:energybal}) in terms
of $f$ the frequency of observed radiation. Equating $2 \pi f$ to $2 \omega$  
(keeping in mind that the frequency of radiation is twice that of the orbital frequency), 
we have that $\omega = \pi f$. 
Making this substitution in eq (\ref{eq:energybal}) we obtain
\begin{equation}
{\cal M} = \frac{c^3}{G} \left( \frac{ 1}{3 \alpha} \pi^{-8/3} f^{-11/3} \frac{df}{dt} \right)^{3/5}.
\label{eq:ligoequation}
\end{equation}
Setting $\alpha = 32/5$ Eq (\ref{eq:ligoequation}) precisely matches the only equation in the LIGO discovery paper. \cite{ligo}
Eq (\ref{eq:ligoequation}) shows that as the black
holes spiral inward the frequency of the emitted radiation increases rapidly. This is the famous
chirp. To make this more explicit we can integrate eq (\ref{eq:ligoequation}) to show that \cite{ftnt:chirp}
\begin{equation}
\frac{1}{f_1^{8/3}} - \frac{1}{f_2^{8/3}} = 8 \alpha \pi^{8/3} \frac{G^{5/3} {\cal M}^{5/3} }{ c^5} \tau.
\label{eq:chirp}
\end{equation}
Here $f_1$ is the initial frequency and $f_2$ is the frequency after a time $\tau$. Eq (\ref{eq:chirp})
shows that the frequency rises rapidly from the value $f_1$ and diverges in a finite amount of time
$\tau_\infty$. Of course the divergence is spurious and the rise in frequency is eventually interrupted by the
coalescence of the black holes as we will see below. 

{\em Estimating the chirp mass.} The bottom panel in fig 1 shows the variation of the frequency of
the observed gravitational radiation as a function of time. The frequency indeed rises rapidly as 
predicted by eq (\ref{eq:chirp}). Assuming $f_2 \gg f_1$ we may neglect the second term on the right
hand side of eq (\ref{eq:chirp}) to obtain
\begin{equation}
{\cal M} =\frac{c^3}{G f_1}\frac{1}{(8 \alpha f_1 \tau)^{3/5}\pi^{8/5}}
\label{eq:chirpestimate}
\end{equation}
Inspection of the bottom left panel of figure 1 shows that $f_1 \approx 42$ Hz at time 0.35 s and that
the frequency essentially diverges at time 0.43 s. Hence $\tau \approx 0.08$ s. Substituting these
values of $f_1$ and $\tau$ in eq (\ref{eq:chirpestimate}) yields ${\cal M} = 35$ solar masses,
very close to the value of 30 solar masses obtained by the LIGO collaboration. In keeping 
with the spirit of a back of the envelope estimate we do not attempt to estimate the uncertainty
in our result. However if we pick other reasonable values of $f_1$ and $\tau$
from the data we always find values of ${\cal M}$ close to 30 $M_{{\odot}}$. 

{\em Bound on the total mass.} The chirp mass gives an idea of the scale of the system 
but it does not by itself reveal the individual masses of the two black holes. If the two black
hole masses are equal then it follows from the definition of the chirp mass that the total
mass of the pair is $(4)^{3/5} {\cal M}$. For ${\cal M} = 30 M_{{\odot}}$ this amounts to 
a total mass of $70 M_{{\odot}}$ or $35 M_{{\odot}}$ per black hole. 
More generally one can show that if the chirp mass is ${\cal M}$ then the total mass of the 
pair has to be greater than $(4)^{3/5} {\cal M}$. To show this assume that the $m = \xi M_{{\rm tot}}$
and $M = (1-\xi) M_{{\rm tot}}$. Here the fraction $\xi$ lies in the range $0 <\xi < 1$.
From the definition of the chirp mass it follows 
\begin{equation}
M_{{\rm tot}} = \frac{{\cal M}}{ [ \xi (1 - \xi) ]^{3/5} }.
\label{eq:fractions}
\end{equation}
Over the unit interval the quantity $\xi(1-\xi)$ is maximum at $ \xi = 1/2$ which rigorously 
justifies the lower bound on $M_{{\rm tot}}$ noted above. 

{\em Schwarzschild radius; Event horizon.} It is a standard exercise in introductory physics
to show that at a distance $r$ from the center of a spherical body of mass $M$ the
escape velocity is $\sqrt{2 G M/r}$ (assuming that $r$ is greater than the radius of the 
spherical body). Setting the escape velocity equal to the speed of light $c$ we obtain the
Schwarzschild radius $R_s$ of the body given by
\begin{equation}
R_s = \frac{2 G M}{c^2}.
\label{eq:schwarzschild}
\end{equation}
If the spherical body is sufficiently dense that it is smaller than $R_s$ then it has an event
horizon and is a black hole. The event horizon is a sphere of radius $R_s$ that surrounds
the black hole. Particles that travel slower than the speed of light can escape the black hole
only if they remain outside the event horizon. Although our analysis is based on Newtonian
gravity the same conclusion emerges from classical general relativity: only particles
that remain outside the event horizon can escape the black hole and the 
radius of the event horizon is given by eq (\ref{eq:schwarzschild}). Furthermore 
in classical general relativity a particle that is inside the event horizon can never
emerge outside, a restriction not present in Newtonian gravity.  \cite{weinberg}

{\em Merger and chirp cutoff.} The black holes will begin to coalesce once their separation is equal to
the sum of their Schwarzschild radii. In other words
\begin{equation}
R+ r = \frac{2 G}{c^2} ( M + m ).
\label{eq:coalesce}
\end{equation}
At this separation according to eq (\ref{eq:keplerlaw}) the angular frequency of the orbiting
black holes is 
\begin{equation}
\omega_c = \frac{1}{\sqrt{8}} \frac{c^3}{G(M+m)}.
\label{eq:coalescefrequency}
\end{equation}
Hence we estimate that the highest frequency attained by the chirp is $f_c = \omega_c/\pi$. 

{\em Estimating the black hole masses.} It follows from eq (\ref{eq:coalescefrequency}) that
\begin{equation}
M_{{\rm tot}} = M+m = \frac{1}{\pi \sqrt{8}} \frac{c^3}{G f_c}.
\label{eq:mtot}
\end{equation}
Inspection of the bottom left panel of figure 1 suggests the value $f_c = 300$ Hz 
which yields $M_{{\rm tot}} = 76 M_{{\odot}}$ , in remarkable agreement with 
$M_{{\rm tot}} = 65 M_{{\odot}}$ obtained by the LIGO collaboration.
In order to determine the individual masses of the black holes we use 
${\cal M} = 30 M_{{\odot}}$ and $M_{{\rm tot}} = 76 M_{{\odot}}$. 
Eq (\ref{eq:fractions}) then yields $\xi = 0.7$ implying that the individual
black hole masses are 53 and 23 solar masses respectively (in fair 
agreement with the values of 36 and 29 solar masses obtained by the LIGO collaboration). 

{\em Estimating the total energy radiated.} Perhaps the most awe-inspiring fact about GW150914 is the
staggering amount of energy emitted in the form of gravitational radiation. The LIGO collaboration
determined that the energy equivalent of 3 solar masses was emitted
in just a tenth of a second. Making use of Einstein's formula
$E = mc^2$ that is approximately $5 \times 10^{47}$ J. By way of comparison 
the sun emits $4 \times 10^{25}$ J in the form of electromagnetic radiation in a tenth of a second;
over its entire lifespan of several billion years the sun will radiate less than  one percent of its mass. 
We can estimate the total gravitational energy radiated by using eq (\ref{eq:energy}) for the energy
of the orbiting black holes. \cite{ftnt:energycaveat}
For the purpose of this estimate we assume that the final separation of the
black holes is the sum of their Schwarzschild radii and that the initial separation is much larger and 
may be taken to be essentially infinite. Substituting eq (\ref{eq:coalesce}) in eq (\ref{eq:energy}) 
yields 
\begin{equation}
\frac{1}{4} \left( \frac{Mm}{M+m} \right) c^2
\label{eq:erad}
\end{equation}
as the estimate of the total amount of gravitational wave energy radiated. Setting $M = 36 M_\odot$ and $m = 29 M_\odot$ yields 4$M_\odot$ as the estimated energy radiated 
in good agreement with the value of 3$M_\odot$
determined by LIGO. Eq (\ref{eq:erad}) also reveals that for a fixed total mass $M+m$, 
the radiated energy is maximized when the \
merging black hole masses are equal and diminishes greatly if one of the merging objects is much lighter
than the other. 

{\em Estimating the initial separation.} Another quantity we can estimate once the total mass of the 
two black holes is known is the 
separation of the black holes when the gravitational radiation first
becomes detectable. 
From the bottom left panel of figure 1 we see
that the frequency of gravitational radiation is approximately 45 Hz when the in-spiral process 
first becomes observable.
Substituting this frequency in eq (\ref{eq:keplerlaw}) and taking the total mass of the two black holes
to be 65 $M_\odot$ 
reveals that the initial separation of the black holes was about 800 km. The combined
Schwarzschild radii of the two black holes on the other hand are approximately 100 km.

{\em Gravitational wave amplitude.} Gravitational waves alternately stretch and compress the space through which
they propagate. The amplitude of a gravitational wave denoted $h$ is the fractional amount by which the wave
stretches or compresses space or the distance between the two ends of an interferometer arm; 
$h$ is a dimensionless
quantity. Assuming that the intensity of a gravitational wave is proportional to the square of the amplitude, 
it is a simple exercise in dimensional analysis to show that the intensity is given by
\begin{equation}
{\cal I}_{{\rm rad}} = \beta h^2 \frac{f^2 c^3}{G}.
\label{eq:intensity}
\end{equation}
The intensity is the energy flow per unit time per unit area normal to the direction of propagation
and $\beta$ is a dimensionless constant of order unity. A full analysis based on linearized general relativity shows that $\beta = \pi/2$. \cite{weinberg}

{\em Estimating the distance.} The intensity of gravitational radiation at a distance $R$ from in-spiraling binary 
black holes is directional but on average falls off with distance as ${\cal I}_{{\rm rad}} = P_{{\rm rad}}/4 \pi R^2$. 
Using this inverse square law, eq (\ref{eq:binarypower}) and eq (\ref{eq:intensity}) we can then
determine $R$ in terms of the observed $h$ and $f$ as
\begin{equation}
R = \frac{4}{\sqrt{5}} \pi^{2/3} \frac{G^{5/3}}{c^4} \frac{1}{h} f^{2/3} \frac{m M}{(m+M)^{1/3}}.
\label{eq:distance}
\end{equation}
From the second panel on the left of figure 1 we see that
$h \approx 10^{-21}$ for $f \approx 250$ Hz. Using these 
values and $m = 29$ and $M = 36$ solar masses we determine the distance of the GW150914 to be
1.7 billion light years in good agreement with the 1.3 billion light years determined by LIGO. 
The agreement is surprisingly good since the same caveats apply to eq (\ref{eq:distance}) as to 
eq (\ref{eq:erad}) and in addition we have ignored the directionality of quadrupole radiation. 

{\em Directional information.}
With just two detectors in operation it was not possible to pinpoint the exact direction of the source.
The signal arrived at the Louisiana detector before the one in Washington by 7 milliseconds. 
Given the approximate latitude and longitude of the detectors 
 one can deduce that the distance
between the detectors is 3000 km or 10 ms at the speed of light. 
If the direction of propagation of the gravitational waves was parallel to the displacement
between the two sites the delay would be exactly 10 ms. On the other hand if the direction
of propagation was perpendicular to the displacement vector the signal would arrive simultaneously
at the two detectors. The delay of 7 ms implies that the direction of propagation makes an angle of
45$^\circ$ with the displacement vector (see problem). GW150914 therefore lies on a circle on the sky
that subtends an angle of 45$^\circ$. This is as far as one can go with a single delay time
but the LIGO collaboration used the entire time dependence of both signals to further pinpoint
particular arcs along this circle that are more likely
to have been the location of the source (see figure 4 of ref \onlinecite{ligoprops}). 
Directional information is important because it allows study of
the same source by other more traditional channels such as gamma ray and neutrino astronomy.
In the future it is expected that the two LIGO detectors at Livingston and Hanford will be joined by a 
third detector that will allow 
more precise location of future sources. 


\section{Discussion}

The above Newtonian analysis gives surprisingly good agreement with the parameters 
obtained by the fully relativistic treatment used by LIGO. In addition to the numerous simplifications 
explicitly stated above, we also ignore polarization of the gravitational radiation and the spin of the
black holes\footnote{The spin of the black holes leads to additional velocity-dependent
forces between the black holes during in-spiral. Minute forces of this kind that act on satellites and
gyroscopes in Earth orbit due to the rotation of the Earth have been experimentally measured. For
binary black holes undergoing in-spiral these forces are much more significant due to the larger masses and 
relativistic velocities involved.
For simplicity we have not included these refinements in our
analysis.}. Incorporation of these effects and other refinements is certainly possible but is contrary to 
the spirit of the ballpark estimates that we wished to present. Because of the approximations made
our estimates should be correct only to order of magnitude (although serendipitously in many instances
our estimates are within a factor of two of the best fit obtained by LIGO).

Prior to LIGO the best evidence for gravitational waves came from observations of binary
pulsars.\cite{taylornobel, taylor} Whereas LIGO is able to detect the actual distortion of space caused by the
passage of gravitational waves, binary pulsars only provide indirect evidence for gravitational waves. Furthermore
it is estimated that the best studied Hulse-Taylor binary pulsar will take 300 million years to 
coalesce (see problem 8 in appendix B). Thus for the foreseeable future the binary pulsar provides access
only to the in-spiral phase whereas LIGO's binary black hole has provided a view of the strong gravity 
physics of coalescence and ring-down. 

Finally we briefly discuss physics after the merger.
During the ring-down phase the signal from the merged black hole resembles the transients of a
under-damped
harmonic oscillator familiar from introductory physics (see figure 1). 
This transient corresponds to the longest lived ``quasi-normal'' mode of the
black hole space-time. The damping rate and ringing frequency of the quasi-normal mode are 
determined by the mass and spin of the quiescent black hole that forms after the quasi-normal
modes have died away.
Thus the spins of the initial black holes can be determined using the in-spiral data and
the spin of the final merged object using the ring-down data. By verifying that the initial and final spins
are consistent with each other using a numerical analysis of the merger the LIGO team were able for the first time 
to test General Relativity in the hitherto inaccessible strong field regime, 
another significant outcome of their discovery.\cite{viewpoints, ligogr}

%

{\em Note added.} After completion of this work we learnt of a new e-print from the LIGO-VIRGO collaboration
that covers some of the same ground as the present manuscript.\cite{ligoped} We thank Andrew Matas, Laleh Sadeghian
and Madeleine Wade for bringing ref \cite{ligoped} to our attention and Ofek Birnholtz and Alex Nielsen for a
helpful correspondence. 

\appendix

\section{Quadrupole Tensor}

\label{sec:quadrupole}

The purpose of this section is to introduce the quadrupole tensor by analogy to the moment
of inertia tensor. This section thus requires some familiarity with the moment of inertia tensor,
which is not typically included in an introductory physics course but is accessible at the level of
\cite{feynmanvol2}. 
Consider a system of $n$ particles with mass $m_i$ and position $(x_i, y_i, z_i)$ with $i = 1, 2, 3, \ldots, n$.
The moment of inertia tensor has components
\begin{equation}
I_{xx} = \sum_{i=1}^n m_i (y_i^2 + z_i^2); \hspace{2mm}
I_{xy} = - \sum_{i=1}^n  m_i x_i y_i;
\label{eq:mominertia}
\end{equation}
and the form of the other diagonal and off-diagonal components can be written down by analogy.\cite{feynmanvol2}
The quadrupole tensor is very similar. The off-diagonal components differ only in sign, e.g., $Q_{xy} = - I_{xy}$. The 
diagonal components have the form $Q_{xx} = - I_{xx} + \frac{1}{3} \overline{I}$ and similarly for $Q_{yy}$ and 
$Q_{zz}$. Here $\overline{I} = I_{xx} + I_{yy} + I_{zz}$. 
In terms of the quadrupole tensor the power of gravitational radiation
emitted is given by
\begin{equation}
P_{{\rm rad}} = \frac{1}{5} \frac{G}{c^5} \sum_{\alpha,\beta = x, y, z} 
\left( \frac{d^3 Q_{\alpha \beta}}{d t^3} \right)^2.
\label{eq:quadruplepower}
\end{equation}
Now suppose for simplicity that the rotating system lies in the $x$-$y$ plane and rotates about the
$z$-axis. After a half rotation the components of the the quadrupole tensor will have returned to their
original values. As far as the quadrupole moment goes the radiating system is back to its original state after
just half a rotation; hence the frequency of the radiation is twice the frequency of rotation. 

\section{Problems} 

\label{sec:probs}

Problems 1, 2 and 3 provide additional information about the LIGO discovery; problem 4 is on
a different application of the two body circular orbit discussed in Section II: and problems 5, 6 and 7
treat electromagnetic radiation and radiation reaction effects that are analogous to the 
gravitational effects analyzed in the main body of the paper.
Problem 8 analyzes the 
gravitational radiation due to the in-spiral of the Hulse-Taylor binary pulsar.
The problems vary in level of difficulty. 

{\em Problem 1. What's in a name?} (a) The discovery event is designated GW150914. Explain why.
{\em Hint:} Re-read the first paragraph of the Introduction. (b) The second black hole
merger event observed by LIGO was designated GW151226. \cite{ligo2} What can you deduce from the name of the event?

{\em Problem 2. Sensitivity of LIGO.} The length of an arm of LIGO's interferometer is 4 km.
By how much is this distance changed due to a gravitational wave with amplitude $h=10^{-21}$?
Give your answer in m. Compare to the radius of a proton (approximately 0.9$\times 10^{-15}$ m). 

{\em Problem 3. Delay and direction.} (a) Using the latitude and longitude information for the
two detectors verify that the distance between the two detectors is 3000 km.
(b) Suppose that the direction of propagation makes an angle $\theta$ with the displacement from the Livingston detector to
the Hanford detector. Show that delay in the arrival at Hanford is given by $d \cos \theta / c$
where $d$ is the distance between the two detectors. 

{\em Problem 4. Discovery of the first extra solar planet.} A star of mass $M$ and a planet of mass
$m$ orbit their center of mass in circular orbits of radius $R$ and $r$ respectively at a common
frequency $\omega$. The orbits
are in the $x$-$y$ plane and the center of mass is at the origin. The earth lies on the negative $y$-axis
many light years distant. The planet cannot be seen from Earth but the $y$-component of the velocity
of the star can be measured spectroscopically and will evidently oscillate at the frequency
$\omega$ with an amplitude $R \omega$. For the star 51 Pegasus the velocity was observed to oscillate
with a period of 4.23 days and an amplitude of 60 m/s leading to the discovery the first extra solar planet,
51 Pegasi B.c\cite{51peg}
Spectroscopically the star resembles the sun and
hence it is reasonable to assume $M = M_{{\odot}} = 2.0 \times 10^{30}$ kg. (a) Use eq (\ref{eq:keplerlaw})
and the measured period to determine $R+r$ the distance between the star and planet. You may neglect
the mass of the planet. Give your answer in astronomical units. (1 A.U. = 1.5 $\times 10^{11}$ m). 
(b) Use the measured amplitude and period to determine $R$. Then combine with your result of part
(a) to determine $m$, the mass of the planet. Compare to the mass of Jupiter (2 $\times 10^{27}$ kg). 

{\em Answer to Problem 4.} (a) 0.05 A.U. (b) $R = 3.5 \times 10^6$ m. $m = 0.9 \times 10^{27}$ kg. 

{\em Problem 5. Electromagnetic dipole radiation.} 
(a) {\em Dimensional analysis.} 
In the S.I. system it is most convenient to express the
units of all quantities in terms of kg, m, s and C. 
For example Coulomb's law reveals that
the units of $\epsilon_0$ are C$^2$\;s$^2$/kg\;m$^3$. 
Determine the units
of the electric dipole moment and the magnetic dipole moment. 
\newline
(b) {\em Electric dipole radiation.}
A rotating electric
dipole moment will produce radiation. 
Determine the dependence of the power radiated $P_{{\rm electric}}$ on 
the magnitude of the dipole moment, $p$, the angular
frequency of rotation, $\omega$, and the basic electromagnetic constants $\epsilon_0$ and $c$
using dimensional analysis. 
(c\hspace{0mm}) {\em Magnetic dipole radiation.} 
Determine the dependence of the power radiated by an rotating magnetic dipole, $P_{{\rm mag}}$, on the
magnitude of the dipole moment, $d$, the angular frequency of rotation, $\omega$, and the electromagnetic
constants $\epsilon_0$ and $c$ using dimensional analysis. 

{\em Answer 5.} (a) Electric dipole moment: C m. Magnetic dipole moment: C m$^2$/s. (b) $P_{{\rm electric}} = (1/6\pi) 
(\omega^4 p^2/\epsilon_0 c^3)$. 
This result is called Larmor's formula. 
(c\hspace{0mm}) $P_{{\rm mag}} = (1/6 \pi \epsilon_0) (\omega^4 d^2/\epsilon_0 c^5)$. 
In both (b) and (c\hspace{0mm}) the pre-factor of $1/6 \pi$ cannot be determined by dimensional analysis; a
full derivation based on Maxwell's equations is necessary. 

{\em Problem 6. Spin down of the Crab Nebula Pulsar.} (a) Model the pulsar as a sphere with moment of 
inertia $I$ and a magnetic dipole $d$ that rotates at an angular frequency $\omega$. Assuming that the dominant
mechanism for the pulsar to lose rotational kinetic energy is via magnetic dipole radiation show that the pulsar
slows down according to the equation $d \omega/dt = - \omega^3 \tau$ where the time scale $\tau = d^2/\epsilon_0 I
c^5$. (b) Integrate the result of part (a) to show that $\omega^{-2} - \omega_0^{-2} = 2 t \tau$ where $\omega_0$ is the
angular frequency of the pulsar at time $t=0$ when it was initially created, $\omega$ is the angular frequency at time $t$.
Assuming that $\omega_0 \gg \omega$ it follows that the age of the pulsar is $t \approx 1/2 \omega^2 \tau$.
(c\hspace{0mm}) The Crab Nebular pulsar was discovered in the 1960s. At that time astronomers measured
$\omega = 200$ s$^-1$ and $d\omega/dt = 2.6 \times 10^{-9}$ s$^{-2}$. Use these data to determine
$\tau$ and the age in years 
of the Crab Nebular pulsar. The pulsar is believed to have been born from the supernova recorded
by Chinese astronomers in the year 1054. \cite{ostriker}

{\em Problem 7. Classical lifetime of Bohr atom.} According to the Bohr model the electron orbits
an essentially stationary proton in an orbit of radius $a_B = 0.53 {\rm \AA}$. Show that according to
classical electromagnetism the electron should spiral inward and merge with the proton in a time
$\tau = (1/4) (4 \pi \epsilon_0 m/e^2)^2 c^3 a_B^3$ where $m$ is the mass of an electron. 
Determine $\tau$ in seconds. For simplicity you may use non-relativistic mechanics throughout. 

{\em Hint for Problem 7.} Calculate the total kinetic and potential energy for a circular orbit of 
radius $r$. Note that as the electron orbits the electric dipole moment of the atom rotates leading
to electric dipole radiation. Use the result of problem 5(b) and energy conservation to show that
the electron spirals inward. 

{\em Problem 8. Gravitational radiation from binary pulsar.} The Hulse-Taylor binary pulsar consists
of two neutron stars, one a pulsar.\cite{taylornobel} By timing the pulsar it has been discovered
that the orbital period is 7.75 hours and the two stars have a mass of approximately 1.4 $M_{{\odot}}$ each.
(a) Assuming a circular orbit use eq (\ref{eq:energybal}) to estimate how much the period decreases per year. 
(b) Use eq (\ref{eq:chirp}) to estimate the time for the binary pulsar to coalesce. 

{\em Answer 8.} (a) 10 $\mu$s. (b) 1 billion years. Discussion: The orbit is actually highly elongated
with an eccentricity of 0.62. Because of the regularity of the pulses and the exquisite precision with which
their times are measured the masses of the neutron stars are actually known to five significant figures and
the eccentricity to seven. Taking the eccentricity into account \cite{peters} it is found that the period 
actually decreases 76.5 $\mu$s per year which agrees with the observed value to within one percent.\cite{taylor} 
The time to coalesce is 300 million years when eccentricity is taken into account.

\vskip.15in
{\bf Figure 1}\\
Summary of LIGO data (reproduced from the discovery paper). \cite{ligo} The top left panel 
shows the strain $h$ observed by the Hanford detector as a function of time; the top right panel shows
the data for the Livingston detector with the Hanford data time-shifted, inverted and overlaid to show the excellent match
between the two detectors.
The data have been band pass filtered to lie in the 35-350 Hz band of maximum detector sensitivity; spectral line
noise features in the detectors within this band have also been filtered. The second row shows
a fit to the data using sine-Gaussian wavelets (light gray) and a different waveform reconstruction (dark gray). 
Also shown in color are the signals obtained from numerical relativity using the best fit parameters to the data.
The third row shows for both detectors 
the residuals obtained by subtracting the numerical relativity curve from the filtered data
in the first row. The fourth row gives a time-frequency representation of the data and shows the
signal frequency increasing in time.

\begin{thebibliography}{99}
\bibitem{ftnt:selfenergy} For the duration
of the in-spiral it is acceptable to treat the black holes as particles of fixed mass and the only relevant
gravitational potential energy is the mutual potential energy of the two black holes, not their individual self-energy.
\bibitem{ftnt:energy} The reader will notice that the total energy is one half the potential energy. 
This is an example of the virial theorem in classical mechanics, but students of introductory
physics can derive the result by adding up all the contributions to the
orbital energy. The potential energy is obviously $-GMm/(r+R)$. The kinetic energy is
given by $\frac{1}{2} I \omega^2$. Making use of eq (\ref{eq:keplerlaw}) for $\omega$ and
eq (\ref{eq:moment}) for the moment of inertia leads quickly to the result quoted in eq (\ref{eq:energy}) for the 
total energy.
\bibitem{ftnt:dipolerad} By way of comparison,
electromagnetic radiation is predominantly produced by the changing electric or magnetic dipole
moments. The quadrupole component matters only if the dipole moments of the radiating system
vanish by symmetry. See problem 5 in appendix B for more on electromagnetic dipole radiation.
\bibitem{ftnt:chirp} To obtain eq (\ref{eq:chirp}) first write eq (\ref{eq:ligoequation}) in the form $f^{-11/3} \; d f = 3 \alpha \pi^{8/3} ({\cal M} G)^{5/3} (1/c^5) \; d t$ and integrate both sides.
Take the lower limit on the time integral $0$ and the upper limit as $\tau$. 
The corresponding limits on the $f$ integral are $f_1$ and $f_2$ respectively, yielding eq (\ref{eq:chirp}).
\bibitem{ftnt:energycaveat} By so doing we are approximating the total energy radiated by the energy
radiated during the in-spiral process alone. Although the peak power is radiated during the merger, 
inspection of fig 1 suggests that the power radiated during in-spiral is indeed comparable to the total
power radiated during the entire process, thereby justifying the estimate in eq (\ref{eq:erad}).
\end{thebibliography}
\end{document}